\documentstyle[prd,aps,
preprint, tighten,
multicol
]{revtex}

\begin{document}

\newcommand{\TeV}{\,{\rm TeV}}
\newcommand{\GeV}{\,{\rm GeV}}
\newcommand{\MeV}{\,{\rm MeV}}
\newcommand{\keV}{\,{\rm keV}}
\newcommand{\eV}{\,{\rm eV}}
\def\bea{\begin{eqnarray}}
\def\eea{\end{eqnarray}}
\def\be{\begin{equation}}
\def\ee{\end{equation}}

\setcounter{page}{1}
\draft

\preprint{KIAS-P99002, hep-ph/9901220}

\title{Axino-Neutrino Mixing in Gauge-Mediated Supersymmetry Breaking Models}
\author{Eung Jin Chun}
\address{ Korea Institute for Advanced Study, \\
          207-43 Cheongryangri-dong, Dongdaemun-gu,
          Seoul 130-012, Korea}
\maketitle

\begin{abstract}
When the strong CP problem is solved by spontaneous breaking of 
an anomalous global symmetry  in theories with gauge-mediated supersymmetry 
breaking, the pseudo Goldstone fermion (the axino)  is a good candidate
of a light sterile neutrino.  Its mixing with neutrinos relevant 
for current neutrino experiments can arise in the presence
of R-parity violation. The realistic four neutrino mass matrix is 
obtained when the see-saw mechanism is brought in, 
and an ansatz for the right-handed neutrino mass is constructed.
\end{abstract}


Current neutrino experiments observing the deficits in 
the atmospheric \cite{ATM} and solar \cite{SOL}
neutrino fluxes possibly hint at the existence 
of an $SU(3)_c\times SU(2)_L \times U(1)_Y$ singlet fermion that mixes with
the ordinary neutrinos.  More excitingly, the reconciliation of 
the above neutrino data with the candidate events for 
the $\nu_\mu \to \nu_e$ oscillation \cite{LSND} requires the
existence of such a singlet fermion (called a sterile neutrino \cite{VM})
with the mass $\lesssim \cal{O}$(1) eV.
Since the mass of a singlet state cannot be protected by the 
gauge symmetry of the standard model,  the introduction of a sterile 
neutrino must come with a theoretical justification.
In view of this situation, there have been many attempts
to seek for the origin of a light singlet fermion 
with the required properties \cite{STERI,II}.
In this letter, we point out that a sterile neutrino can arise naturally
in a well-motivated extension of the standard model, namely, 
in the supersymmetric standard model (SSM) incorporating the Peccei-Quinn 
(PQ) mechanism for the resolution of the strong CP 
problem \cite{KIM}, and the mechanism of supersymmetry breaking through
gauge mediation \cite{GMSB}.  

\medskip

Most attractive solution to the strong CP problem would be the PQ mechanism
which introduces a QCD-anomalous global symmetry 
(called the PQ symmetry) spontaneously broken
at a high scale $f_a \approx  10^{10}-10^{12}$ GeV, 
and thus predicts a pseudo Goldstone boson (the axion, $a$) \cite{KIM}.
In supersymmetric theories, the fermionic partner of the axion (the
axino, $\tilde{a}$) exists and would be massless 
if supersymmetry is conserved.
But supersymmetry is broken in reality and there would be a large mass
splitting between the axion and the axino, which depends on the mechanism
of supersymmetry breaking.
In the context of supergravity where supersymmetry breaking is mediated at 
the Planck scale  $M_P$, the axino
mass is generically of the order of the gravitino mass, 
$m_{3/2} \sim 10^2$ GeV \cite{LUK}, which characterizes the supersymmetry
breaking scale of the SSM sector.
However, the axino can be very light if supersymmetry breaking occurs 
below the scale of the PQ symmetry breaking, as in 
theories with the gauge-mediated supersymmetry breaking (GMSB) \cite{GMSB}.
The GMSB models are usually composed of three sectors: the SSM, messenger and
hidden sector.  The messenger sector contains extra vector-like quarks and 
leptons which conveys supersymmetry breaking from the hidden sector to the
SSM sector \cite{DINE}.

\medskip

In order to estimate the axino mass in the framework of GMSB,
it is useful to invoke the non-linearly realized Lagrangian 
for the axion superfield.  
Below the PQ symmetry breaking scale $f_a$,
the couplings of the axion superfield $\Phi$ can be rotated away from the
superpotential and are encoded in the K\"ahler potential as follows:
\bea \label{KPhi}
 K &=& \sum_I C_I^\dagger C_I + \Phi^\dagger \Phi + 
    \sum_I {x_I \over f_a} (\Phi^\dagger + \Phi)C_I^\dagger C_I  \nonumber\\
 &+& \mbox{higher order terms in $f_a$} 
\eea
where $C_I$ is a superfield in the SSM, messenger, or hidden sector,
and $x_I$ is its PQ charge.  Upon supersymmetry breaking, the axino gets 
the mass from the third term in Eq.~(\ref{KPhi}),
\be
 m_{\tilde{a}I} \approx  x_I  {F_I \over f_a } 
\ee
where $F_I$ is the F-term of the field $C_I$.
If there is a massless fermion charged under the PQ symmetry,
the axino would have a Dirac mass of order $F_I/f_a$ \cite{NT}\footnote{
Note that the late-decaying particle scenario for the structure formation
with the axino in the MeV region and the gravitino in the eV region 
\cite{CKK} can be realized in this case for $F\sim (10^5 \GeV)^2$.}.
It is however expected that there is  no massless mode 
[except neutrinos, see below] in the theory, and each component of the 
superfield $C_I$ has the mass of order $M_I \sim \sqrt{F_I}$ unless one
introduces extra symmetries to ensure the existence of massless modes.
Then, the axino mass is see-saw reduced to have the  majorana mass:
\be
 m_{\tilde{a}} \approx {x_I^2 F_I^2 \over M_I f_a^2} 
 \sim x_I^2{M_I^3 \over f_a^2} \,.
\ee
Now one can think of three possible scenarios implementing the PQ symmetry:
PQ symmetry acting (i) only on the SSM sector, (ii) on the messenger sector,
(iii) on the hidden sector.  In each case, $\sqrt{F_I}$ is of order of
(i) $10^2-10^3$ GeV, (ii) $10^4-10^5$ GeV, (iii) $10^5-10^6$ GeV.
In the last case, we restricted ourselves to have a light gravitino 
$m_{3/2} \lesssim 1$ keV which evades overclosure of the universe
if no entropy dumping occurs after supersymmetry breaking.
Then we find the following ranges of the axino mass:
\be \label{maxino}
m_{\tilde{a}} \sim \cases{ 
(10^{-9}-10^{-6})\eV \left(10^{12} \GeV \over f_a \right)^2\quad\mbox{(i)}\cr
(10^{-3}-1)\eV \left(10^{12} \GeV \over f_a \right)^2  \quad \mbox{(ii)} \cr
(1-10^3)\eV \left(10^{12} \GeV \over f_a \right)^2 \quad \mbox{  (iii)} \cr }
\ee
Interestingly, the axino mass in the case (i) or (ii) falls into the
region relevant for the current neutrino experiments.  
Our next question is then how the axino-neutrino mixing arises.

\medskip

The mixing of the axino with neutrinos can come from the axion 
coupling to the lepton doublet $L$, $x_\nu (\Phi^\dagger+\Phi) 
L^\dagger L$, which yields the mixing mass,
\be
 m_{\tilde{a}\nu} \approx x_\nu {F_\nu \over f_a} 
\ee
where $x_\nu$ is the PQ charge of the lepton doublet.
It is crucial for us to observe that 
{\it nonzero} $F_\nu$ can arise when R-parity is {\it not} imposed.
To calculate the size of $F_\nu$, we work in the basis where the 
K\"ahler potential takes the canonical form.
That is, $L^\dagger H_1+{\rm h.c.}$ is rotated away in the K\"ahler potential,
and  the superpotential of the SSM allows for the bilinear terms,
\be \label{Wbi}
W=\mu H_1 H_2 + \epsilon_i \mu L_i H_2 \,. 
\ee
where the dimensionless parameter $\epsilon_i$ measures 
the amount of R-parity violation.
Due to the $\epsilon$-term  in Eq.~(\ref{Wbi}), one has 
$F_\nu=\epsilon_i \mu v \sin\beta$ where $v=174$ GeV and 
$\tan\beta=\langle H_2 \rangle / \langle H_1 \rangle$, and therefore,
\be \label{maxnu}
m_{\tilde{a}\nu_i} \sim  10^{-4} \eV  
     \left(\epsilon_i \sin\beta \over 2\times10^{-6}\right)
     \left(10^{12} \GeV \over f_a\right)
     \left( \mu \over 300 \GeV \right)\,.
\ee
Since R-parity and lepton number violating bilinear operators (\ref{Wbi})
are introduced,  we have to take into account the so-called tree-level 
neutrino mass \cite{HS}. The tree mass  arises due to the misalignment of
the sneutrino vaccum expectation values with the $\epsilon_i$ terms, which   
vanishes at the mediation scale $M_m$ of supersymmetry breaking,
but is generated at the weak scale through renomalization group (RG) evolution
of supersymmetry breaking parameters.  This tree mass takes the form of
$m_{\nu}^{\rm tree} \propto \epsilon_i \epsilon_j$, and 
its size (dominated by one component $\epsilon_i$) is given by 
\cite{JA,CHC},
\bea \label{mtree}
 m^{\rm tree}_{\nu_i} &\approx& 1 \eV 
       \left( a_i \over 3\times 10^{-6} \right)^2
       \left( M_Z \over M_{1/2} \right) \nonumber \\
 \mbox{where}\quad  a_i &\sim &  \epsilon_i \sin\beta 
           \left(\mu A_b \over m_{\tilde{l}}^2 \right)
           \left( {3 h_b^2 \over 8\pi^2}\ln{M_m \over m_{\tilde{l}} } \right)
\eea
where $M_Z, M_{1/2}$  and $m_{\tilde{l}}$ are the Z-boson, gaugino and slepton
mass, respectively,  and $h_b, A_b$ are the $b$-quark Yukawa coupling and
corresponding trilinear soft-parameter, respectively.
In Eq.~(\ref{mtree}), the term proportional to $h_b^2 \ln(M_m/ m_{\tilde{l}})$ 
characterizes the size of the RG-induced misalignment.
Taking $\tan\beta=1, \mu A_b = m_{\tilde{l}}^2$,
and $M_m =10^3 m_{\tilde{l}}$ as reference values, 
one finds
\be \label{mnu}
m^{\rm tree}_{\nu_i} \sim 10^{-4} \eV
       \left( \epsilon_i \over 10^{-6} \right)^2
       \left( M_Z \over M_{1/2} \right) 
\ee
which grows roughly as $\tan^4\beta$ in large $\tan\beta$ region.

\medskip

Based on Eqs.~(\ref{maxino}), (\ref{maxnu}) and (\ref{mnu}), 
we can make the following observations:
\begin{itemize}
\item
The just-so solution of the solar neutrino problem with large mixing and 
$\Delta m^2_{\rm sol} \approx 0.7\times10^{-10} \eV^2$ \cite{SOL} implies 
$m^{\rm tree}_{\nu_e}, m_{\tilde{a}} < m_{\tilde{a} \nu_e} \sim 10^{-5} \eV$,
and could be realized in the case (i). 
For this, we need $f_a \gtrsim 10^{11}$ GeV and $\epsilon_1 \sim 10^{-7}$.
\item
The small mixing MSW solution requiring $\theta_{\rm sol} \approx 
4\times 10^{-4}$ and  $\Delta m^2_{\rm sol} \approx 5\times 10^{-6} \eV^2$
\cite{SOL} can be realized for the case (i) or (ii) if 
$ m^{\rm tree}_{\nu_e} < m_{\tilde{a}} \approx 2\times 10^{-3}$ eV and 
$m_{\tilde{a}\nu_e} \approx 10^{-4} \eV $.  For this, one needs 
$f_a\sim 10^{10}$ GeV and $\epsilon_i \sim 10^{-8}$ in the case (i); or 
$f_a\approx 10^{12}$ GeV and $\epsilon_i \sim 10^{-6}$ in the case (ii).
\item
The $\nu_\mu \to \tilde{a}$ explanation of the atmospheric neutrino
oscillation \cite{CKL} requiring nearly maximal mixing  \cite{ATM}
is realized 
if $m_{\tilde{a}\nu_e} > m_{\tilde{a}}, m^{\rm tree}_{\nu_\mu}$ and 
$\Delta m^2_{\rm atm} \approx 2 m_{\tilde{a}\nu_e}(m_{\tilde{a}}+ 
m^{\rm tree}_{\nu_\mu}) \sim 3\times 10^{-3} \eV^2$.
The best region of parameters for this is 
$\epsilon_2 \sim 10^{-5}$ and $f_a \sim 10^{10}$ GeV prefering the case
(i).
\end{itemize}
Note that a low $\tan\beta$ is preferred to suppress 
$m^{\rm tree}_{\nu_i}$ in all of the above cases.

\medskip

Having seen that the $\nu_{e,\mu}-\tilde{a}$ mixing can arise
to explain the solar or atmospheric neutrino problem,  let us now discuss 
how all the experimental data \cite{ATM,SOL,LSND}
can be accommodated in our scheme.  
Recall that there exist only two patterns of neutrino mass-squared differences
compatible with the results of all the experiments. 
Namely, four neutrino masses are divided into two pairs
of almost degenerate masses separated by a gap of
$\sqrt{\Delta m^2_{\rm LSND} } \sim 1 \, \mathrm{eV} $
as indicated by the result of the LSND experiments \cite{LSND}, and follow 
either of the following patterns \cite{BGG}:
\begin{eqnarray}
&& \mbox{(A)}
\qquad \underbrace{ \overbrace{m_1 < m_2}^{\mathrm{atm}}
                \quad \ll \quad
                    \overbrace{m_3 < m_4}^{\mathrm{solar}}
                  }_{\mathrm{LSND}} \,,
\nonumber     \\[-2mm]
&& \label{AB} \nonumber\\[-2mm] 
&& \mbox{(B)}
\qquad \underbrace{ \overbrace{m_1 < m_2}^{\mathrm{solar}}
 \quad \ll \quad
                    \overbrace{m_3 < m_4}^{\mathrm{atm}}
                  }_{\mathrm{LSND}} \;.
\nonumber
\end{eqnarray}
In (A), $\Delta{m}^{2}_{21}$ is relevant for the explanation
of the atmospheric neutrino anomaly and $\Delta{m}^{2}_{43}$
is relevant for the suppression of solar $\nu_e$'s.
In (B), the roles of $\Delta{m}^{2}_{21}$
and $\Delta{m}^{2}_{43}$ are interchanged.

In our scheme, the required degeneracy of $m_3$ and $m_4$
could be a consequence of the quasi Dirac structure: $m_{\tilde{a}\nu_i}
\gg m_{\tilde{a}}, m_{\nu_i}^{\rm tree}$.  To have $m_4, m_3 \approx
m_{\tilde{a}\nu_i} \sim 1 \eV$, we need a large value of $\epsilon_i$ 
$ \sim 10^{-4}$ e.g. for $f_a =10^{10}$ GeV.  But this makes  the 
tree mass too large ($m^{\rm tree}_{\nu_i} \sim 1 \eV$) 
to accommodate $\Delta m^2_{\rm sol}$ or $\Delta m^2_{\rm atm}$ for (A) or (B),
respectively.  For the pattern (B), the atmospheric neutrino oscillation 
could also be explained by the $\nu_\mu-\nu_\tau$ degeneracy with 
the splitting $m_4-m_3 \sim 10^{-3} \eV$.
However, when the ordinary neutrino mass comes from R-parity violation, 
the neutrino mass takes  generically the 
hierarchical structure since the tree mass 
(of the form $m^{\rm tree}_{ij} \propto \epsilon_i \epsilon_j$) 
is rank-one. Even though this structure can be changed due to
the one-loop contribution through the squark and slepton exchanges 
\cite{HS,CHC}, one needs fine-tuning to achieve the required 
degeneracy: $(m_4-m_3)/m_3 \sim 10^{-3}$ \cite{CHC}. 

The simplest way to get the  realistic neutrino mass matrix would be 
to introduce heavy right-handed neutrinos $N$ whose masses are 
related naturally to the PQ scale $f_a \sim 10^{12} \GeV$.  
For this purpose, let us
assign the following $U(1)$ PQ charges to the fields:
\be
\begin{array}{ccccccccc}
 H_1 & H_2 & L & N & \phi & \phi' & \sigma & \sigma' & Y \\
 1   & 1   & 2 & -3& -1   &  1    &   6    & -6      & 0 \\
\end{array}
\ee
Then the PQ symmetry allows for the superpotential 
$W=W_{N} + W_{PQ}$ where  
\bea
 W_N &=& {h_\mu \over M_P}H_1 H_2 \phi^2 + {h_i \over M_P^2} L_i H_2 \phi^3
        + {m^D_i \over \langle H_2 \rangle } L_i H_2 N_i
        + {M_{ij} \over 2 \langle \sigma \rangle} N_i N_j \sigma \nonumber \\
 W_{PQ} &=& (\phi \phi'+ \sigma\sigma' + M_1^2)Y+ M_2 Y^2 + 
        {h_y \over 3} Y^3 + {1\over 6} \phi^6\sigma + 
         {1\over 6} {\phi'}^6\sigma' 
\eea
where $M_{ij}, M_1, M_2 \sim f_a$.
Minimization of the scalar  potential coming from $W_{PQ}$ leads to the 
supersymmetric minimum with the vaccum expectation values
$\langle \phi \rangle, \langle \phi' \rangle, 
\langle \sigma \rangle, \langle \sigma' \rangle \sim f_a$ and 
$\langle Y\rangle \sim 0$ satisfying the conditions
$\langle \phi/\phi' \rangle^6 = \langle \sigma'/\sigma \rangle$ and 
$\langle \phi\phi' \rangle = 6 \langle \sigma \sigma' \rangle $.  
Note that the superpotential $W_N$ provides
a solution to the $\mu$ problem \cite{MU}, and  generates the 
right value of $\epsilon_i$ for our purpose, that is,
\be
 \mu \sim {h_\mu f_a^2 \over M_P}\sim 10^2 \GeV \,, \quad 
 \epsilon_i \sim   {h_i f_a \over h_\mu M_P} \sim 10^{-6}  \,.
\ee
Now the usual arbitrariness comes in the determination of the 
Dirac neutrino mass $m^D_i$ and the right-handed
neutrino mass $M_{ij}$.   Guided by the unification spirit,
let us first take the Dirac masses, $m^D_1\sim 1 \MeV$, $m^D_2\sim 1 \GeV$ and 
$m^D_3\sim 100 \GeV$  following the hierarchical structure of up-type quarks. 
Then, the see-saw suppressed neutrino mass matrix 
$m^{\rm 3x3}_{ij}=- m^D_i m^D_j M^{-1}_{ij}$ takes the form,
\be \label{SS}
 m^{\rm 3x3} \sim 1 \eV 
      \pmatrix{ 10^{-8}c_{11} & 10^{-5}c_{12} & 10^{-3}c_{13} \cr
                10^{-5}c_{12} & 10^{-2}c_{22} & c_{23}       \cr
                10^{-3}c_{13} & c_{23}       & 10^2c_{33}  \cr }
\ee
where $c_{ij} = M^{-1}_{ij}/ M^{-1}_{23}$ and $m^D_2m^D_3M^{-1}_{23} \sim
1 \eV$.  Note that the matrix (\ref{SS}) is close to 
the neutrino mass matrix ansatz \cite{BWW}:
\be \label{ANm}
 m^{\rm 3x3} \sim 1 \eV 
 \pmatrix{ 0  & 0 & 0.01  \cr
           0 &  10^{-3} & 1       \cr
           0.01 & 1 &  \lesssim 10^{-3}  \cr }
\ee
which explains the atmospheric neutrino and LSND data,
except that (\ref{SS}) has a too large component, $m^{\rm 3x3}_{33}$.
Restricted by minimality condition and requiring $c_{33}=0$, 
the ansatz (\ref{ANm}) can be translated to the corresponding 
one for the right-handed neutrino mass matrix:
\be \label{ANM}
 M_{ij} = \pmatrix{ A & B & C \cr B & B^2/A & 0 \cr C & 0 & 0 \cr }  \,.
\ee 
The parameters $A,B,C$ in Eq.~(\ref{ANM}) are determined from the observational
quantities as follows:
\bea
 \sqrt{\Delta m^2_{\rm LSND} } &\approx& 
         {A m^D_2m^D_3 \over BC}\nonumber\\
 \theta_{\rm LSND} &\approx& {B\over A} {m^D_1 \over m^D_2} \\
 {\Delta m^2_{\rm atm} \over \Delta m^2_{\rm LSND}} &\approx& {C\over B}
  {m^D_2 \over m^D_3}  \nonumber
\eea
from which one finds $A,C \sim 0.1 B$, and $B^2/A \sim 10B$ for 
$B \approx 10^{11} \GeV$. 
 
\medskip

In conclusion, it has been shown that the axino, which is predicted by the
PQ mechanism in supersymmetric theories, can be naturally light and
mix with ordinary neutrinos  to explain the solar or atmospheric neutrino 
problem in the context of gauge mediated supersymmetry breaking.  
The lightness is ensured by the small supersymmetry breaking scale $\sqrt{F}
\lesssim 10^5$ GeV (far below the PQ symmetry breaking scale $f_a$)
and the mixing is induced due to the presence of
the R-parity violating bilinear term $\epsilon_i \mu L_iH_2$.
The $\nu_e-\tilde{a}$ mixing can arise to explain the solar neutrino
deficit when the parameters are in the ranges: $\epsilon_1 \sim 10^{-6}$ and 
$f_a \sim 10^{12}$ GeV 
for which the axion can provide cold dark matter of the universe.   
To account for the atmospheric neutrino oscillation ($\nu_\mu \to \tilde{a}$),
larger $\epsilon$ and smaller $f_a$ are preferred: $\epsilon_2 \sim 10^{-5},
f_a \sim 10^{10} \GeV$. 

With R-parity violation alone, it is hard to find the four 
neutrino mass matrix which accommodates all the neutrino data. 
It has been therefore attempted to obtain a realistic four neutrino 
oscillation scheme relying on the see-saw mechanism
combined with R-parity violation.  In this case, only the pattern (B) with 
the $\nu_e \to \tilde{a}$ solar  and $\nu_\mu\to \nu_\tau$ atmospheric 
oscillations can be realized.  A toy model has been built
in a way that the right values of $\mu$ and $\epsilon$ are generated naturally
through the PQ symmetry selection rule.
In the unification scheme where the Dirac neutrino mass follows the 
hierarchical structure of up-type quarks, an ansatz for the
right-handed neutrino mass matrix has been constructed to
reproduce the required light neutrino mass matrix.

%
\end{document}